\newcommand{\minpoint}{\mbox{$'\mskip-4.7mu.\mskip0.8mu$}}
\documentstyle[emulateapj]{article}

\lefthead{Ajiki et al.}
\righthead{Possible Superwind Galaxy at $z$ = 5.69}

\begin{document}

\submitted{Accepted for publication in the Astrophysical Journal}

\title{A NEW HIGH-REDSHIFT LYMAN$\alpha$ EMITTER:
       POSSIBLE SUPERWIND GALAXY AT $z$ = 5.69 
         \altaffilmark{1,} \altaffilmark{2}}

\author{Masaru Ajiki        \altaffilmark{3},
          Yoshiaki Taniguchi  \altaffilmark{3},
          Takashi Murayama    \altaffilmark{3},
          Tohru Nagao         \altaffilmark{3},
          Sylvain Veilleux    \altaffilmark{4},
          Yasuhiro Shioya     \altaffilmark{3},
          Shinobu S. Fujita   \altaffilmark{3},
          Yuko Kakazu         \altaffilmark{5},
          Yutaka Komiyama     \altaffilmark{6},
          Sadanori Okamura    \altaffilmark{7,8},
          David B. Sanders    \altaffilmark{5},
          Shinki Oyabu        \altaffilmark{9},
          Kimiaki Kawara      \altaffilmark{9},
          Youichi Ohyama      \altaffilmark{6},
          Masanori Iye        \altaffilmark{10},
          Nobunari Kashikawa  \altaffilmark{10},
          Michitoshi Yoshida  \altaffilmark{11},
          Toshiyuki Sasaki    \altaffilmark{6},
          George Kosugi       \altaffilmark{6},
          Kentaro Aoki        \altaffilmark{9},
          Tadafumi Takata     \altaffilmark{6},
          Yoshihiko Saito     \altaffilmark{10},
          Koji S. Kawabata    \altaffilmark{10},
          Kazuhiro Sekiguchi  \altaffilmark{6},
          Kiichi Okita        \altaffilmark{6},
          Yasuhiro Shimizu    \altaffilmark{11},
          Motoko Inata        \altaffilmark{9},
          Noboru Ebizuka      \altaffilmark{12},
          Tomohiko Ozawa      \altaffilmark{13},
          Yasushi Yadoumaru    \altaffilmark{13},
          Hiroko Taguchi      \altaffilmark{14},
          Hiroyasu Ando       \altaffilmark{6},
          Tetsuo Nishimura    \altaffilmark{6},
          Masahiko Hayashi    \altaffilmark{6},
          Ryusuke Ogasawara   \altaffilmark{6}, \&
          Shin-ichi Ichikawa  \altaffilmark{10}
         }

\altaffiltext{1}{Based on data collected at the
         Subaru Telescope, which is operated by
         the National Astronomical Observatory of Japan.}
\altaffiltext{2}{Based on data collected at
          the W. M. Keck Observatory}
\altaffiltext{3}{Astronomical Institute, Graduate School of Science,
          Tohoku University, Aramaki, Aoba, Sendai 980-8578, Japan}
\altaffiltext{4}{Department of Astronomy, University of Maryland,
          College Park, MD 20742, USA}
\altaffiltext{5}{Institute for Astronomy, University of Hawaii,
          2680 Woodlawn Drive, Honolulu, HI 96822, USA}
\altaffiltext{6}{Subaru Telescope, National Astronomical Observatory,
          650 N. A'ohoku Place, Hilo, HI 96720, USA}
\altaffiltext{7}{Department of Astronomy, Graduate School of Science,
          University of Tokyo, Tokyo 113-0033, Japan}
\altaffiltext{8}{Research Center for the Early Universe, School of Science,
          University of Tokyo, Tokyo 113-0033, Japan}
\altaffiltext{9}{Institute of Astronomy, Graduate School of Science,
          University of Tokyo, 2-21-1 Osawa, Mitaka, Tokyo 181-0015, Japan}
\altaffiltext{10}{National Astronomical Observatory of Japan,
          2-21-1 Osawa, Mitaka, Tokyo 181-8588, Japan}
\altaffiltext{11}{Okayama Astrophysical Observatory,
          National Astronomical Observatory,
          Kamogata-cho, Asakuchi-gun, Okayama 719-0232, Japan}
\altaffiltext{12}{Communications Research Laboratory, 4-2-1 Nukui-Kitamachi,
          Koganei, Tokyo 184-8795, Japan}
\altaffiltext{13}{Misato Observatory, 180 Matsugamine, Misato-cho, Amakusa-gun,
          Wakayama, 640-1366, Japan}
\altaffiltext{14}{Department of Astronomy and Earth Sciences, Tokyo Gakugei
          University, 4-1-1 Nukui-Kitamachi, Koganei, Tokyo 184-8501, Japan}

\begin{abstract}
During the course of our deep optical imaging survey for Ly$\alpha$
emitters at $z \approx 5.7$ in the field around the $z=5.74$ quasar
SDSSp J104433.04-012502.2, we have found a candidate strong
emission-line source.  Follow-up optical spectroscopy shows that the
emission line profile of this object is asymmetric, showing excess
red-wing emission. These properties are consistent with an
identification of Ly$\alpha$ emission at a redshift of $z=5.687 \pm
0.002$.  The observed broad line width, $\Delta v_{\rm FWHM} \simeq
340$ km s$^{-1}$ and excess red-wing emission also suggest that this
object hosts a galactic superwind.
\end{abstract}


\section{INTRODUCTION}

Recent progress in deep optical imaging with 8-10 m class telescopes
has enabled new searches for star-forming galaxies beyond redshift
5. In particular, imaging surveys using narrow-passband filters have
proved to be a particularly efficient way to find such galaxies (Hu \&
McMahon 1996; Cowie \& Hu 1998; Steidel et al. 2000; Kudritzki et
al. 2000; Hu et al. 2002).  Indeed the most
distant Ly$\alpha$ emitter known to date is HCM 6A at $z=6.56$ (Hu et
al. 2002), and more than a dozen Ly$\alpha$ emitters beyond $z=5$ have
been discovered (Dey et al. 1998; Spinrad et al. 1998; Weymann et
al. 1998; Hu et al. 1998, 1999, 2002; Dawson et al. 2001, 2002; Ellis
et al. 2001), most by using this
technique. 

One interesting object is J123649.2+621539 at $z=5.190$
which was found serendipitously in the HDF-North flanking fields
(Dawson et al. 2002). Its Ly$\alpha$ emission-line profile shows a
sharp blue cutoff and broad red wing emission, both of which are often
observed in star-forming systems with prominent wind outflows. These
features are also expected from radiative transfer in an expanding
envelope. Therefore, Dawson et al. (2002) suggested that the
Ly$\alpha$ profile of J123649.2+621539 is consistent with a superwind
with a velocity of $\sim$ 300 km s$^{-1}$.
Galactic superwinds are now considered to be one of the key issues for
understanding the interaction and evolution of both galaxies and
intergalactic matter (e.g. Heckman 1999; Taniguchi \& Shioya 2000). 
In order to improve our knowledge of galactic superwinds at high
redshift, a large sample of superwind candidates at $z > 3$ is
needed. During the course of our new search for Ly$\alpha$ emitters at
$z \approx 5.7$, we have found a candidate superwind galaxy at $z =
5.69$.  In this {\it Letter}, we report its observed properties.  We
adopt a flat universe with $\Omega_{\rm matter} = 0.3$,
$\Omega_{\Lambda} = 0.7$, and $h=0.7$ where $h = H_0/($100 km s$^{-1}$
Mpc$^{-1}$) throughout this {\it Letter}.

\section{OPTICAL DEEP IMAGING}

We have carried out a very deep optical imaging survey for faint Ly$\alpha$
emitters in the field surrounding the quasar SDSSp
J104433.04$-$012502.2 at redshift 5.74\footnote{
The discovery redshift was $z=5.8$ (Fan et al. 2000).
Since, however, the subsequent optical spectroscopic observations
suggested a bit lower redshift; $z=5.73$ (Djorgovski et al. 2001)
and $z=5.745$ (Goodrich et al. 2001), we adopt $z=5.74$ in this Letter.}
(Fan et al. 2000; Djorgovski et
al. 2001; Goodrich et al. 2001), using the prime-focus wide-field
camera, Suprime-Cam (Miyazaki et al. 1998) on the 8.2 m Subaru
Telescope (Kaifu 1998) on Mauna Kea.  Suprime-Cam consists of ten
2k$\times$4k CCD chips and provides a very wide field of view --
$34^\prime \times 27^\prime$ (0.2 arcsec pixel$^{-1}$).  In this
survey, we used the narrow-passband filter, NB816, centered on 8160
\AA ~ with a passband of $\Delta\lambda_{\rm FWHM} = 120$ \AA; the
central wavelength corresponds to a redshift of 5.72 for Ly$\alpha$
emission. We also used broad-passband filters, B, R$_{\rm C}$, I$_{\rm
C}$, and z$^\prime$. A summary of the imaging observations is given in
Table 1. All observations were done under photometric conditions, and
the seeing was between 0$\farcs$7 and 1$\farcs$3 arcsec during the
entire run.  Photometric and spectrophotometric standard stars used in
the flux calibration are SA101 for the $B$, $R_{\rm C}$, and $I_{\rm
C}$ data, and GD 50, GD 108 (Oke 1990), and PG 1034+001 (Massey et
al. 1996) for the NB816 data. The $z^\prime$ data were calibrated by
using the magnitude of SDSSp J104433.04$-$012502.2 (Fan et al. 2000);
since any quasar is a potentially variable object, the photometric 
calibration of the $z^\prime$ data may be more unreliable than
those of the other-band data.

To avoid delays in obtaining follow-up spectroscopy, we analyzed only
two of the CCD chips, one of which included the quasar SDSSp
J104433.04-012502.2.  The individual CCD data were reduced and
combined using IRAF and the mosaic-CCD data reduction software
developed by Yagi et al. (2002).  The total size of the reduced
subfield is $11\minpoint67 \times 11\minpoint67$, corresponding to a
total solid angle of $\approx$ 136 arcmin$^{2}$. The volume probed by
the NB816 imaging has (co-moving) transverse dimensions of 27.56
$h_{0.7}^{-1}$ Mpc $\times 27.56 h_{0.7}^{-1}$ Mpc, and the FWHM
half--power points of the filter correspond to a co-moving depth along
the line of sight of 44.34 $h_{0.7}^{-1}$ Mpc ($z_{\rm min} \approx
5.663$ and $z_{\rm max} \approx 5.762$; note that the transmission
curve of our NB816 filter has a Gaussian-like shape). Therefore, a
total volume of 33,700 $h_{0.7}^{-3}$ Mpc$^{3}$ is probed in our NB816
image.

Source detection and photometry were performed using SExtractor
version 2.2.1 (Bertin \& Arnouts 1996).  Our detection limit (a
3$\sigma$ detection with a $2\arcsec$ diameter aperture) for each band
is summarized in Table 1.  For source detection in the NB816 image, we
used a criterion that a source must be an object with at least a
13-pixel connection at the 5$\sigma$ level. [Note: Given the pixel
resolution of 0.2 arcsec pixel$^{-1}$, an extended source observed in
$\sim$ 1 arcsec seeing must be observed as a source with more than
a 13-pixel connection.]  Adopting the criterion for the NB816 excess,
$I_{\rm C} - NB816 > 1.0$ mag, we have found a strong emission-line
source located at $\alpha$(J2000.0)=10$^{\rm h}$ 44$^{\rm m}$ 32$^{\rm s}$ and
$\delta$(J2000.0)=$-01^\circ$ 30$^\prime$ 34$^{\prime\prime}$ (hereafter
LAE J1044$-$0130). In this {\it Letter} we report on this source.

The sky position of LAE J1044$-$0130 is shown in Fig. 1.
The optical thumb-nail images of LAE J1044$-$0130 are given in Fig. 2.
As shown in this figure, LAE J1044$-$0130 is clearly seen only in the 
NB816 image; the observed equivalent width is $EW_{\rm obs} > 310$ \AA.
The NB816 image reveals that LAE J1044$-$0130 is spatially extended. 
It is interesting to note that this object shows not a circular shape
but an irregular shape.
The angular diameter is 2.3 arcsec (above the 2$\sigma$ noise level).
The size of the point spread function in the NB816 image is 0.90 arcsec.
Correcting for this spread, we obtain an angular diameter of 2.1
arcsec for the object.  In the cosmology adopted here, this
corresponds to a diameter of $d \simeq 12.4 h_{0.7}^{-1}$ kpc.

\section{OPTICAL SPECTROSCOPY}

In order to investigate the nature of LAE J1044$-$0130, we 
obtained optical spectroscopy.
First, we used the Subaru Faint Object Camera And
Spectrograph (FOCAS; Kashikawa et al. 2000) with 
the lowest spectral resolution grism,
150 lines mm$^{-1}$, blazed at $\lambda$=6500 \AA ~ together
with an order-cut filter SY47 on 2002, March 11 (UT).  The wavelength
coverage was $\sim$ 4900 \AA ~ to 9400 \AA.  The use of an 0.8
arcsec-wide slit gave a spectroscopic resolution of $R \sim 400$ at
8000 \AA. The total integration time was 1800 seconds.  The spectrum
is shown in the upper panel of Fig. 3.  We detected a single emission
line at $\lambda \approx$ 8130 \AA.  This observation was done under
photometric conditions, and we used the spectroscopic standard star,
Feige 34 (Massey et al. 1988) to calibrate the spectrum.  The emission
line flux was calculated to be $(1.5 \pm 0.3) \times 10^{-17}$ ergs
s$^{-1}$ cm$^{-2}$.
A second spectrum  with FOCAS was obtained on 2002, March 13 (UT) in non-photometric
conditions.  This time, we used the 300 lines mm$^{-1}$ grating blazed
at 7500 \AA ~ together with the same order cut filter SY47, giving a
spectral resolution $R \simeq 1000$ with the same 0.8 arcsec-wide
slit.  This setting gave the same wavelength coverage as the lower
resolution observation. Again the total integration time was 1800
seconds.  The spectrum is shown in the lower panel of Fig. 3.  The
emission-line profile marginally shows the sharp cutoff at wavelengths
shortward of the line peak. However, the spectral resolution is
insufficient to confirm this feature.

To clarify the shape of the line profile seen in the FOCAS
observations, additional observations were obtained with Keck/Echelle
Spectrograph and Imager (ESI; Sheinis et al. 2000) in
echellette mode using a 1$\arcsec$ slit, which provided a spectral
resolution of $R \simeq 3400$ at $\lambda$=8000 \AA. Two 1800-second
integrations were obtained in photometric conditions
on 2002, March 15 (UT).  The spectra
were calibrated using the spectroscopic standard stars Feige 34 and HZ
44 (Massey et al. 1988).
The combined spectrum is shown in the middle panel of Fig. 4.  Again
only one emission line, at $\lambda \approx 8130$ \AA\ is found in the
ESI optical spectrum, which covers the wavelength range 4000 \AA\ to
9500 \AA.  The emission-line profile at 8130 \AA\ is confirmed to
slightly asymmetric, showing a cutoff at wavelengths shortward of the
line peak.  

Although a single emission line in the wavelength range 4000 \AA\ to
9500 \AA\ may alternatively be identified as [O {\sc
ii}]$\lambda$3727, the fact that the observed profile appears to be
skewed suggests that this line is in fact Ly$\alpha$ emission
(e.g. Stern et al. 2000; Dawson et al. 2002). 
Therefore, we conclude that the emission line at 8129.3$\pm$3.0 \AA\
must be Ly$\alpha$, which then yields a redshift of 5.687$\pm$0.002.
The rest-frame equivalent width of the putative Ly$\alpha$ emission is
estimated to be $EW_0 > 46$ \AA.  The observed Ly$\alpha$ flux is
$(1.49 \pm 0.33) \times 10^{-17}$ ergs cm$^{-2}$ s$^{-1}$. This
appears consistent with our FOCAS observations made on 2002, March 11.

\section{RESULTS AND DISCUSSION}

\subsection{Star Formation and Superwind Activities}

The observed Ly$\alpha$ flux is $f$(Ly$\alpha$) = $(1.49 \pm 0.33)
\times 10^{-17}$ ergs cm$^{-2}$ s$^{-1}$ based on the ESI spectrum.
Given the cosmology adopted in this {\it Letter}, we obtain an
absolute Ly$\alpha$ luminosity of $L$(Ly$\alpha$) $\simeq (5.3 \pm
1.2) \times 10^{42} ~ h_{0.7}^{-2}$ ergs s$^{-1}$.  This Ly$\alpha$
luminosity is comparable to those of other $z > 5$ galaxies;
1) $3.3 \times 10^{42}$ ergs s$^{-1}$ for HCM 6A at $z=6.56$
     (Hu et al. 2002),
2) $6.1 \times 10^{42}$ ergs s$^{-1}$ for SSA22-HCM1 at $z=5.74$
     (Hu et al. 1999),
3) $3.4 \times 10^{42}$ ergs s$^{-1}$ for HDF 4-473.0 at $z=5.60$
     (Weymann et al. 1998), and
4) $8.5 \times 10^{42}$ ergs s$^{-1}$ for J123649.2+621539.5 at $z=5.19$
     (Dawson et al. 2002).
Note that all the above luminosities are estimated by using the same
cosmology as that used here.

We note that approximately half of the intrinsic Ly$\alpha$
emission from LAE J1044-0130 could be absorbed by intergalactic atomic
hydrogen (e.g., Dawson et al. 2002). In order to reproduce the
observed Ly$\alpha$ emission-line profile a two-component profile fit
was made using the following assumptions:\ 1) the intrinsic Ly$\alpha$
emission line profile is Gaussian, and 2) the optical depth of the
Ly$\alpha$ absorption increases with decreasing wavelength shortward
of the rest-frame Ly$\alpha$ peak. The resulting fit is shown in
Figure 4 (thick curve), which corresponds to the following emission
and absorption line parameters -- 1) Ly$\alpha$ emission: the line
center, $\lambda_{\rm c, em} = 8030.70$ \AA, the line flux, $f_{\rm
em} \simeq 2.42 \times 10^{-17}$ ergs s$^{-1}$ cm$^{-2}$, and the line
width, $FWHM_{\rm em} \simeq 650$ km s$^{-1}$; 2) Ly$\alpha$
absorption: the line center, $\lambda_{\rm c, abs} = 8122.73$ \AA, the
optical depth at the absorption center, $\tau_{\rm abs} \simeq 9.85$,
and the line width, $FWHM_{\rm abs} \simeq 175$ km s$^{-1}$.  This
analysis suggests that the total Ly$\alpha$ emission-line flux amounts
to $1.73 \times 10^{-17}$ ergs s$^{-1}$ cm$^{-2}$.  This is larger by
a factor of 1.16 than the observed flux, giving a total Ly$\alpha$
luminosity of $L$(Ly$\alpha$) $\sim 6.1 \times 10^{42} ~ h_{0.7}^{-2}$
ergs s$^{-1}$.
We then estimate the star formation rate of LAE J1044$-$0130.  Using
the relation $SFR = 9.1 \times 10^{-43} L({\rm Ly}\alpha) ~ M_\odot
{\rm yr}^{-1}$ (Kennicutt 1998; Brocklehurst 1971) and the total
Ly$\alpha$ luminosity, we obtain $SFR = 5.6 \pm 1.1 ~h_{0.7}^{-2} ~
M_\odot$ yr$^{-1}$.  Note that $SFR$ may be overestimated because part
of the Ly$\alpha$ emission may arise from shock-heated gas if the
superwind interpretation is applicable to this object.

Although the signal-to-noise ratio of our ESI spectrum is not high
enough to analyze the profile shape in great detail, the presence of
the excess red-wing emission seems secure (Fig. 4).  The FWHM of the
Ly$\alpha$ emission is measured to be 340$\pm$110 km s$^{-1}$ and the
full width at zero intensity (FWZI) is estimated to be 890$\pm$110 km
s$^{-1}$. These properties are similar to those of the Ly$\alpha$ emitter
at $z=5.190$, J123649.2+621539, found by Dawson et al. (2002).

\subsection{Comments on Possible Association with the Quasar
     SDSSp J104433.04-012502.2 and the Lyman Limit System at $z=5.72$}

The observed redshift of LAE J1044$-$0130, $z=5.687$, is close both to the
quasar redshift, $z=5.74$ (see footnote 15) and to that of a Lyman limit
system (LLS) at $z_{\rm LLS} =5.72$ in the quasar spectrum reported by
Fan et al. (2000).
The redshift difference between LAE J1044$-$0130 and the quasar corresponds
to the velocity difference of $\Delta v \approx 2370$ km s$^{-1}$ and 
that between LAE J1044$-$0130 and the LLS corresponds to $\Delta v \approx 
1476$ km s$^{-1}$. The angular distance between LAE J1044$-$0130 and 
the quasar is approximately 330 arcsec, giving a co-moving separation of
$\sim$ 13 $h_{0.7}^{-1}$ Mpc. This separation seems too large to identify
LAE J1044$-$0130 as a counterpart of the LLS. It seems also unlikely that
that LAE J1044$-$0130 is associated with the large-scale structure in which
the quasar SDSSp J104433.04-012502.2 resides.

\begin{deluxetable}{lcccc}
\tablenum{1}
\tablecaption{Journal of imaging observations}
\tablewidth{0pt}
\tablehead{
\colhead{Band} &
\colhead{Obs. Date (UT)} &
\colhead{$T_{\rm int}$ (sec)\tablenotemark{a}}  &
\colhead{$m_{\rm lim}$(AB)\tablenotemark{b}} &
\colhead{$FWHM_{\rm star}$ (arcsec)\tablenotemark{c}}
}
\startdata
$B$         & 2002 February 17      &  1680 & 27.1 & 1.2 \\
$R_{\rm C}$ & 2002 February 15, 16  &  4800 & 26.8 & 1.4 \\
$I_{\rm C}$ & 2002 February 15, 16  &  3360 & 26.2 & 1.2 \\
$NB816$     & 2002 February 15 - 17 & 36000 & 26.6 & 0.9 \\
$z'$        & 2002 February 15, 16  &  5160 & 25.4 & 1.2 \\
\enddata
\tablenotetext{a}{Total integration time.}
\tablenotetext{b}{The limiting magnitude (3$\sigma$) within a
2$^{\prime\prime}$ aperture.}
\tablenotetext{c}{The full width at half maximum of stellar
objects in the final image}
\end{deluxetable}

\vspace{0.5cm}

We would like to thank both the Subaru and Keck Telescope staff for
their invaluable help. We would also like to thank the referee
for useful comments. This work was financially supported in part by
the Ministry of Education, Culture, Sports, Science, and Technology
(Nos. 10044052, and 10304013).


\begin{figure}
\epsscale{0.50}
\plotone{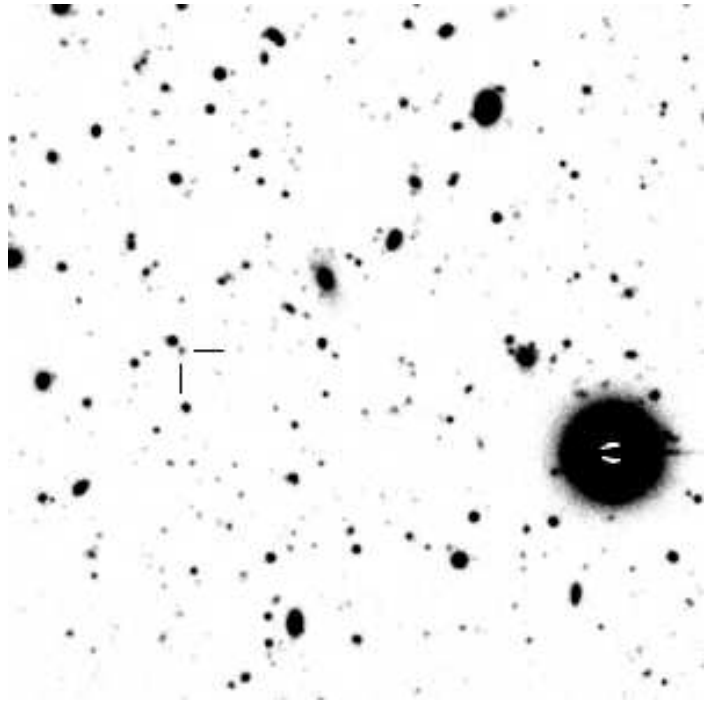}
\caption{ A finding chart of LAE J1044-0130 in our NB816 image.
The field size is 2$^\prime \times 2^\prime$.
North is up and east is to the left.
\label{fig1}}
\end{figure}

\begin{figure}
\epsscale{0.2}
\plotone{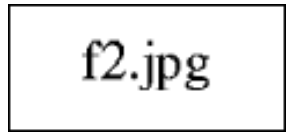}
\caption{ Thumbnail images of LAE J1044-0130 (upper panel)
also displayed as contours (middle panel). The angular size of the circle
in each panel corresponds to 8$\arcsec$.  The lower panel shows
the spectral energy distribution (in magnitudes). From our optical
spectroscopy, we find that  
a galaxy located at 2$\farcs$4 northeast of LAE J1044$-$0130 shows
two emission lines at $\lambda \sim 6710 - 6720$~\AA.
Since these lines can be identified as the [O
{\sc ii}]$\lambda$3727 doublet redshifted to $z=0.802 \pm 0.002$.
This foreground galaxy might enhance
the image of LAE J1044$-$0130 due to gravitational lensing. 
However, since there is no counter image of the LAE J1044$-$0130 
in our deep NB816 
image, the magnification factor might be less than a factor of 2.
\label{fig2}}
\end{figure}

\begin{figure}
\epsscale{0.5}
\plotone{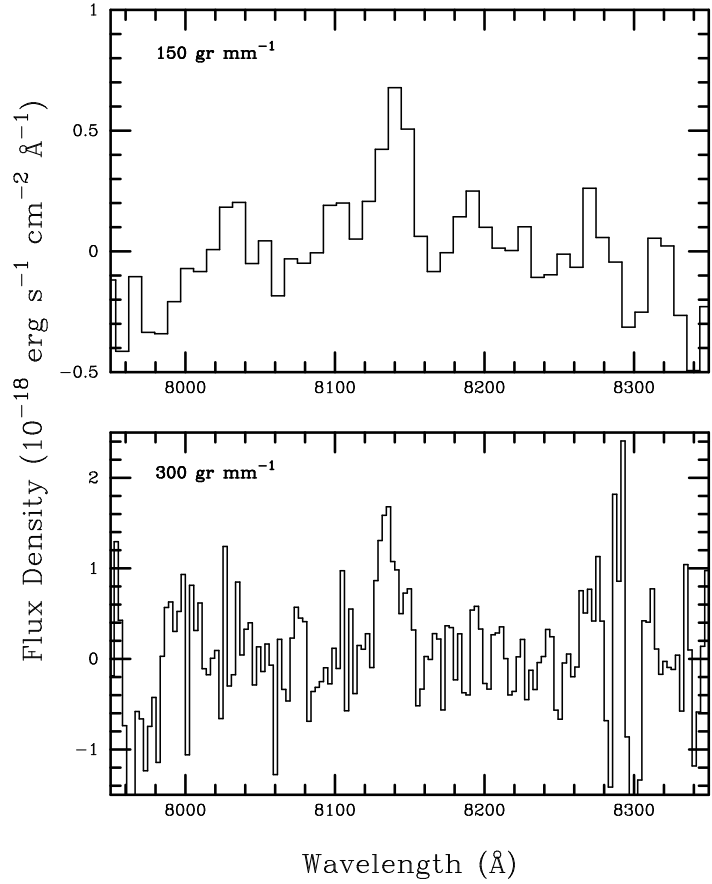}
\caption{ The optical spectra of LAE J1044-0130
obtained with FOCAS on Subaru.  (a) The spectrum shown in the upper
panel ($R \sim 400$) was obtained on 2002, March 11. (b) The spectrum
shown in lower panel ($R \sim 1000$) was obtained on 2002, March 13.
\label{fig3}}
\end{figure}

\begin{figure}
\epsscale{0.2}
\plotone{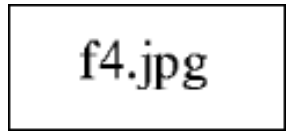}
\caption{ The optical spectrogram (upper panel)
and one-dimensional spectrum (middle panel) of LAE J1044-0130
obtained with ESI on Keck II ($R \sim 3400$); note that five-pixel
binning was applied to the one-dimensional spectrum.  
The model profile fit is 
shown by the thick solid curve (see text).  Sky (OH airglow) emission
lines are shown in the lower panel.
\label{fig4}}
\end{figure}

\end{document}